# Dual-wavelength femtosecond laser-induced low-fluence single-shot damage and ablation of silicon

Alexander V. Bulgakov[*], Juraj Sládek, Jan Hrabovský, Inam Mirza, Wladimir Marine, Nadezhda M. Bulgakova

HiLASE Centre, Institute of Physics ASCR, Za Radnici 828, 25241 Dolni Brezany, Czech Republic

## ABSTRACT

A study of damage and ablation of silicon induced by two individual femtosecond laser pulses of different wavelengths, 1030 and 515 nm, is performed to address the physical mechanisms of dual-wavelength ablation and reveal possibilities for increasing the ablation efficiency. The produced ablation craters and damaged areas are analyzed as a function of time separation between the pulses and are compared with monochromatic pulses of the same total energy. Particular attention is given to low-fluence irradiation regimes when the energy densities in each pulse are below the ablation threshold and thus no shielding of the subsequent pulse by the ablation products occurs. The sequence order of pulses is demonstrated to be essential in bi-color ablation with higher material removal rates when a shorter-wavelength pulse arrives first at the surface. At long delays of 30-100 ps, the dual-wavelength ablation is found to be particularly strong with the formation of deep smooth craters. This is explained by the expansion of a hot liquid layer produced by the first pulse with a drastic decrease in the surface reflectivity at this timescale. The results provide insight into the processes of dual-wavelength laser ablation offering a better control of the energy deposition into material.

**Keywords:** dual-wavelength laser ablation, femtosecond pulses, laser-induced damage, laser craters, silicon, two-temperature model

## 1. INTRODUCTION

There are a number of evidences indicating that ultrashort-laser-based technologies can be substantially improved by using combinations of different wavelengths with higher efficiency and better quality of material processing and nanoparticle generation. Thus, combining infrared and visible laser pulses has been demonstrated to allow for enhancing the ablation efficiency of materials with improved crater morphology [1-4]. The use of bi-chromatic femtosecond pulses can improve the quality of laser-processed surfaces [5], provide control over properties of laser-induced periodic surface structures [6-8], enhance the efficiency of laser energy coupling [4,9] and nanoparticle synthesis [10], and increase the kinetic energy of laser-generated photoelectrons [11].

However, bi-chromatic irradiation regimes are still scantily investigated and mechanisms responsible for enhanced energy coupling and associated improvement of laser processing remain unclear. The governing idea, often suggested for band-gap materials [2,4-6,12,13] is that a shorter wavelength (UV or visible) excites efficiently the valence electrons to the conduction band while a longer wavelength (typically IR) is favorable for heating the generated solid-state plasma. An alternative mechanism assumes the generation of free electrons via photoionization involving different photons [2,9] and thus implies simultaneous action of the pulses for the maximal effect. Furthermore, dual-wavelength drilling can benefit from the formation of dual focus [5]. On the other hand, simulations [12] and experiments [13] indicate that bi-color irradiation is not always advantageous as compared with monochromatic pulses of the same total energy. It is therefore important to reveal conditions when bi-chromatic pulses are particularly efficient.

We recently demonstrated with a silicon target that dual-wavelength femtosecond laser ablation can be very efficient for material removal under certain conditions when an IR pulse arrives at the surface at a fairly long delay of ~100 ps after the visible pulse [4]. In this work, we have continued this study focusing on low-fluence regimes when the energy densities in each pulse are below the ablation thresholds and thus no shielding of the subsequent pulse by the ablation products occurs and the highest benefit from bi-color irradiation is expected. Mechanisms of silicon damage and ablation in these regimes are discussed based on a comparison of the experiments with advanced two-temperature modeling.

*bulgakov@fzu.cz; phone +420 314-007-733; https://www.fzu.cz

## 2. EXPERIMENTAL

Details of the experimental arrangement are described in [4]. In brief, single-crystal silicon (100) was irradiated at normal incidence in the dual-wavelength regime by two individual Gaussian pulses at 1030 and 515 nm wavelengths from a PHAROS laser combined with a HIRO harmonic generator (Light Conversion) with pulse durations 260 and 250 fs, respectively. The beams were combined and focused coaxially by an Al parabolic mirror (100 mm focal length) into the same circular spot on the target surface at normal incidence. Before combining, the 515-nm pulse was delayed by a Michelson interferometer, and the time delay was varied in the range of ±300 ps (the negative delay corresponds to the situation when the visible pulse arrives first). Particular attention was paid to the adjustment of the beam focusing conditions to have identical spot sizes of both beams (effective diameter $D_{eff}$ = 59 μm, $1/e^2$ criterion) to avoid uncertainty in the fluence determination in the dual-wavelength regime. The pulse energies $E_i$ of both beams were adjusted independently using separate attenuators to keep each laser fluence $F_i = 8E_i/\pi D_{eff}^2$ below the corresponding ablation thresholds (see Sec. 4). All the experiments were performed in ambient air under single-shot irradiation conditions.

The zero delay between the 1030- and 515-nm pulses was determined in a special air breakdown experiment in the pump-probe configuration [4]. The laser-produced spots on the surface were examined using an optical microscope in Nomarski mode (Olympus BX43). The morphology of the produced ablation craters and removed material volume were analyzed using a laser-scanning confocal microscope LEXT OLS5000. All the results for the dual-wavelength regime are compared with those obtained under monochromatic irradiation conditions at the same total fluence.

## 3. MODEL

To get insight into the processes of the laser excitation and heating of silicon under dual-wavelength irradiation, we have elaborated the numerical model developed previously [14] for the case of double-pulse action with different laser wavelengths. The model is based on the classical two-temperature model (TTM) and includes the heat-flow equations for the electron and lattice subsystems, the rate equation for the excitation of electrons from the valence band to the conductions band, the optical model describing the dynamic reflection and absorption coefficients, and the equations for the attenuation of laser light in the silicon bulk as described in details in [4]. The photoemission effect was excluded from the model as its role in heating and ablation of silicon is negligible for the considered low laser fluences while considerably increasing the complexity of the modeling [14].

The rate equation describing the excitation of the electrons to the valence band and their recombination takes into account the dual-wavelength-ionization effect as

$$\frac{\partial n_e}{\partial t} = (W_1^{\omega 1} + \frac{1}{2}W_2^{\omega 1}I_{\omega 1})\frac{I_{\omega 1}}{\hbar \omega_1} + W_1^{\omega 2}\frac{I_{\omega 2}}{\hbar \omega_2} + \delta_{imp}n_e - R_{Auger} \qquad (1)$$

Here $n_e$ is the density of the conduction band electrons, $W_1^{\omega 1}$ and $W_2^{\omega 1}$ are one- and two-photon ionization coefficients at 1030 nm; $W_1^{\omega 2}$ is one-photon ionization coefficient at 515 nm, $\delta_{imp}$ is the impact ionization coefficient, and $R_{Auger}$ is the Auger recombination term. The temperature-dependent values for $W_1^{\omega 1}$ and $W_1^{\omega 2}$ were taken from [15,16] whereas the two-photon ionization at 515 nm was neglected [15]. The $W_2^{\omega 1}$ coefficient was taken as 7 cm/GW based on the analysis [4]. The simulations were performed for two Gaussian laser pulses with the time delay $t_{del}$ between them which could be negative or positive. We did not consider time-overlapping pulses in the simulations to avoid their interference that would complicate the modeling because of the unknown probabilities of the mixed-wavelengths excitation.

The dynamically changing optical parameters (reflection and absorption coefficients) were calculated separately for 1030 and 515 nm using the Drude model similar to how it was done in [4,17,18]). Silicon metallization upon melting was assumed with the application of the phase weighting algorithms [19] for the optical constants. The attenuation of the laser beams upon propagation in the sample was described by the following expressions:

$$\frac{\partial}{\partial x}I_{\omega 1}(x,t) = -(W_1^{\omega 1} + W_2^{\omega 1}I_{\omega 1}(x,t) + \alpha_{CB}^{\omega 1}(x,t))I_{\omega 1}(x,t), \qquad (2)$$

$$\frac{\partial}{\partial x}I_{\omega 2}(x,t) = -(W_1^{\omega 2} + \alpha_{CB}^{\omega 2}(x,t))I_{\omega 2}(x,t). \qquad (3)$$

where $\alpha_{CB}^{\omega 1}(x,t)$ and $\alpha_{CB}^{\omega 2}(x,t)$ are the free-electron absorption coefficients. All the parameters used in the calculations and details of the numerical scheme can be found elsewhere [4].

## 4. RESULTS AND DISCUSSION

The femtosecond-laser-induced single-shot damage and ablation thresholds were found previously to be, respectively, 230 and 480 mJ/cm$^2$ for 1030 nm and 130 and 290 mJ/cm$^2$ for 515 nm [4]. In this study, for the dual-wavelength regimes, we kept the intensity of both pulses below the corresponding ablation thresholds but above (or near) the damage thresholds, and thus possible shielding of the second (delayed) pulse by the ablation products produced by the first pulse was excluded. The ultrashort ablation (spallation) of silicon occurs at a ~100 ps timescale [20], therefore, in ablation regimes, the ablation products could affect the coupling of the second pulse as probably took place in [4] where a decrease in the bi-color-laser-induced material removal with increasing fluence was observed. Here we investigate the pure influence of the surface excitation/heating by the first pulse of one wavelength on the coupling to the surface of a second (delayed) pulse of a different wavelength.

Figure 1 shows optical images of spots produced in the dual-wavelength regime with equal energy of 3.0 µJ in 1030-nm and 515-nm laser pulses (corresponding fluence is 220 mJ/cm$^2$) at several time delays between the pulses. Spots produced by monochromatic pulses of double energy (440 mJ/cm$^2$ fluence) are also shown in Fig. 1. As seen, the size and even structure of the spots strongly depend on the sequence order of the pulses and the separation between them. The spots produced at negative and zero delays (i.e., the visible pulse arrives either earlier or simultaneously with the IR pulse, Figs. 1(a)-1(d)) as well as the spot obtained with the 515-nm pulse of double energy (Fig. 1(h)) are fairly large and have a characteristic three-zone structure usually observed at semiconductor surfaces irradiated by individual femtosecond laser pulses in ablation regimes [4,21,22]. The structure includes (1) a central ablation region, (2) a middle annealing region, and (3) an external damage region as illustrated in Fig. 1(h). Interesting that the largest damage areas are observed at near zero delays while the largest ablation areas are produced at fairly long negative delays of around -100 ps. In contrast, spots obtained with positive time delay (IR pulse arrives earlier) are considerably smaller and exhibit no ablation region (Figs. 1(e), 1(f)). Also, the IR pulse at 440 mJ/cm$^2$ is below the silicon ablation threshold (Fig. 1(g)).

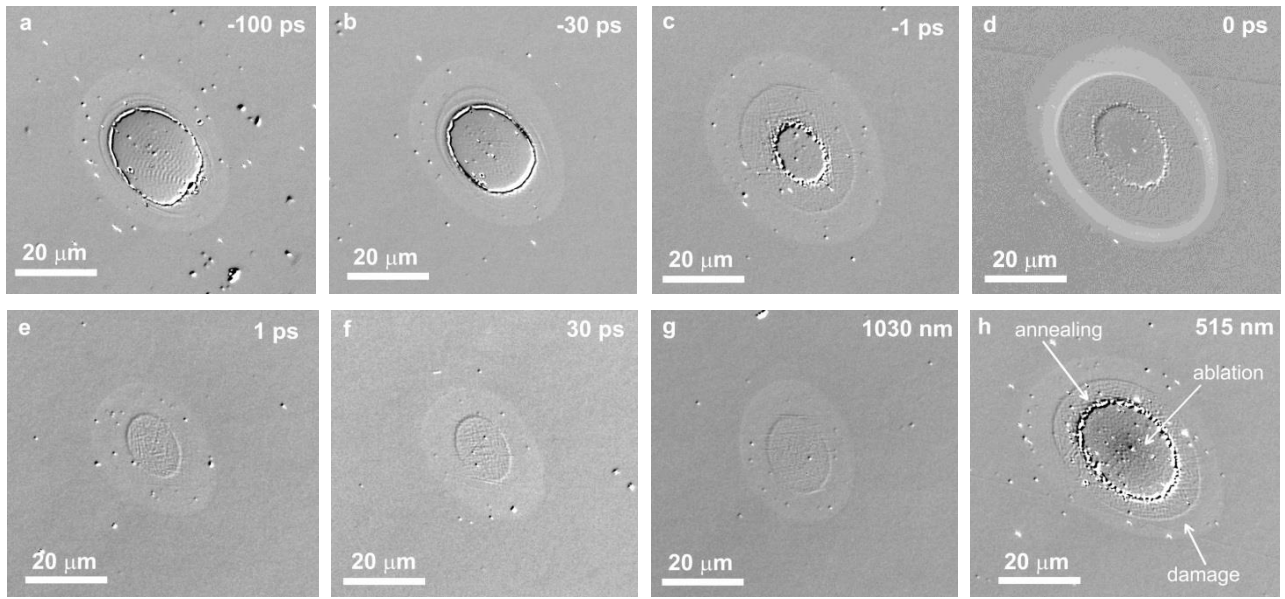

Figure 1. Optical images of spots produced by two laser pulses at 515 and 1030 nm with equal fluences of 220 mJ/cm$^2$ at different time delays between the pulses (*a-f*) and by monochromatic pulses at 1030 nm (*g*) and 515 nm (*h*) at a twofold fluence of 440 mJ/cm$^2$. Three regions are seen within the spots (damage, annealing, and ablation) as indicated in (*h*).

Figure 2(a) shows damage and ablation areas of spots obtained in the dual-wavelength regime with 1030- and 515-nm laser pulses of equal fluences of 220 mJ/cm$^2$ as a function of the delay between the pulses. To reveal whether the bi-color ablation is beneficial or not compared to monochromatic irradiation, it is natural to compare the areas with those obtained with the only 515-nm pulse at the same total fluence since silicon damage and ablation with visible laser pulses is always stronger than those with IR pulses of the same energy under the considered conditions (cf. Figs. 1(g) and (h)). Therefore, the data in Fig. 2(a) are normalized to the corresponding areas of the spots produced by monochromatic irradiation with 515-nm pulses at 440 mJ/cm$^2$. Three remarkable results emerge. First, both dependencies are distinctly asymmetric relative to the zero delay. The damage and ablation of silicon are considerably stronger at negative time delays when the 515-nm pulse comes first, throughout the studied delay range, than those at positive delays. In the latter case, there is no ablation at all. This observation is in line with the theoretical predictions [12] assuming that such a pulse sequence is beneficial for ultrashort laser ablation of band-gap materials since the more energetic visible photons are advantageous for the excitation of electrons from the valence to the conduction band while the following IR pulse is efficiently absorbed by the seed conduction-band electrons via the inverse bremsstrahlung process. Similar asymmetric time-delay dependences of the ultrashort laser ablation rate were observed previously for silicon [2,4] and fused silica [13] in the dual-wavelength regime. Another important result of Fig. 4(a) is that the damaged area in the dual-wavelength regime is maximized at zero delay and is larger than the reference value with a visible pulse only. Finally, the third remarkable feature is the bi-modal time-delay distribution of the ablation area with two peaks, at zero delay and fairly large negative delays of -30 – -100 ps.

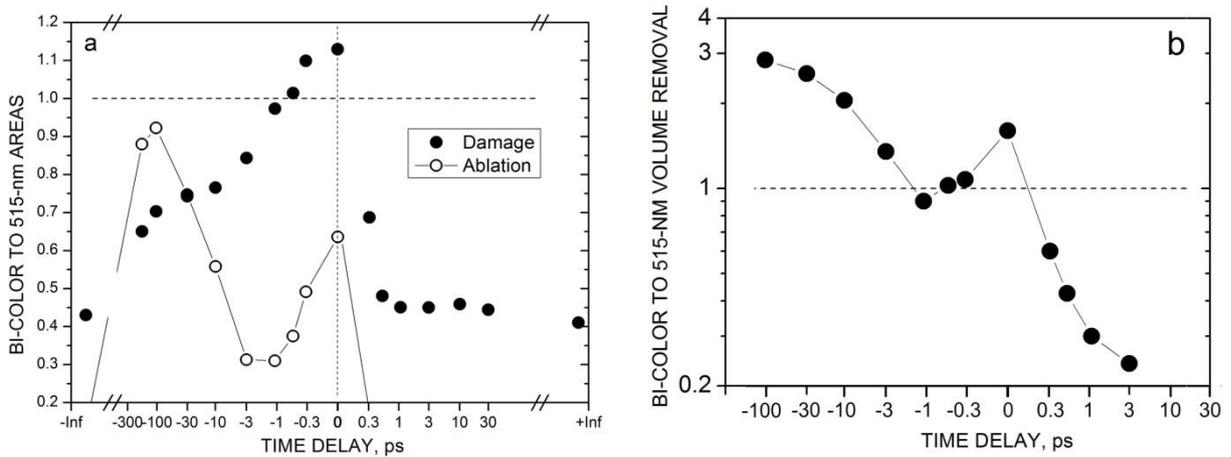

Figure 2. Damaged and ablated areas of spots (*a*) and total volume of craters (b) produced by bi-color irradiation with equal 515-nm and 1030-nm pulse energies as a function of time delay between the pulses for total fluences of 440 J/cm$^2$. The data are normalized to the corresponding values for spots and craters produced by 515-nm pulses at the same fluence. The time axis is given in logarithmic scale to illustrate in more detail the short-delay range. Negative delays correspond to the visible pulse arriving first. The delays ±Inf correspond to long delays of around 10 s.

Figure 3 shows profiles and 3D images of craters produced under dual-wavelength conditions at different time delays and corresponding to the spots shown in Fig. 1. Based on such data, the ablation volume was determined and compared to the material removal in the monochromatic irradiation regime. Figure 2(b) shows the ablation volume obtained from such measurements as a function of the time delay. Again, the data are normalized to the crater volume produced by 515-nm pulses at the same total fluence. Similar to the damage and ablation areas (Fig. 2(a)), a strong asymmetry in the time delay dependence of the crater volume is evident with much larger amounts of material removal at negative delays, when the visible pulse comes first at the surface. The most remarkable feature of the dual-wavelength-produced craters (Fig. 3) is the formation of deep craters at fairly large negative delays of around -30 – -100 ps when the ablation area is maximal (Fig. 2(a)). The removed volume at these delays is up to threefold larger than that for only 515 nm (Fig. 2(b)) even though the ablation area is smaller than the reference one. The craters therewith have a nice nearly cylindrical shape (Figs. 3(a),3b). Therefore, the obtained results demonstrate that a combination of IR and visible femtosecond laser pulses can be beneficial for the ablation of silicon, in terms of both the ablation efficiency and shape of the produced craters, in particular when the IR pulse arrives at the surface at a considerable delay of around 30-100 ps after the visible pulse.

A plausible explanation would be therefore in a relatively slow change of a property of the irradiated surface during a time interval between 10 and 100 ps after the action of the visible pulse. We believe that the observed strong enhancement of the ablation efficiency under dual-wavelength conditions at fairly long delays between the pulses is due to the dynamical change of the reflectivity of the molten silicon layer produced by the first visible pulse. It is well established [23,24] that at low fluences of ultrashort laser pulses, below the ablation threshold but above the melting threshold, the optical reflectivity of the irradiated silicon first increases quickly to the metallic liquid value due to the ultrafast non-thermal melting. The produced hot liquid layer expands thermally resulting in a drastic decrease in the reflectivity and the formation of a dark area at the spot center in a 10-100 ps timescale as was observed by time-resolved microscopy [23,24]. If the second pulse is applied to the surface during the period of its dark state (i.e., at low reflectivity), the pulse energy will be strongly coupled to the target. The fact that the intensity of the first visible pulse is below the ablation threshold is particularly beneficial for the coupling enhancement since otherwise the second IR pulse will be shielded by the products of ablation which is developed for silicon at the same timescale [20]. Of course, a similar effect can be expected also for positive delays when the order of the IR and visible pulses is reversed. However, the reduction in the reflectivity for IR fs-laser-irradiated silicon is considerably lower than in the case of visible irradiation [20] and thus the coupling enhancement effect is much weaker. Therefore, the above scenario appears to be well consistent with our observations. Below we discuss the observed asymmetry of the bi-color irradiation with respect to the sequence order of laser pulses at different wavelengths based on modeling results.

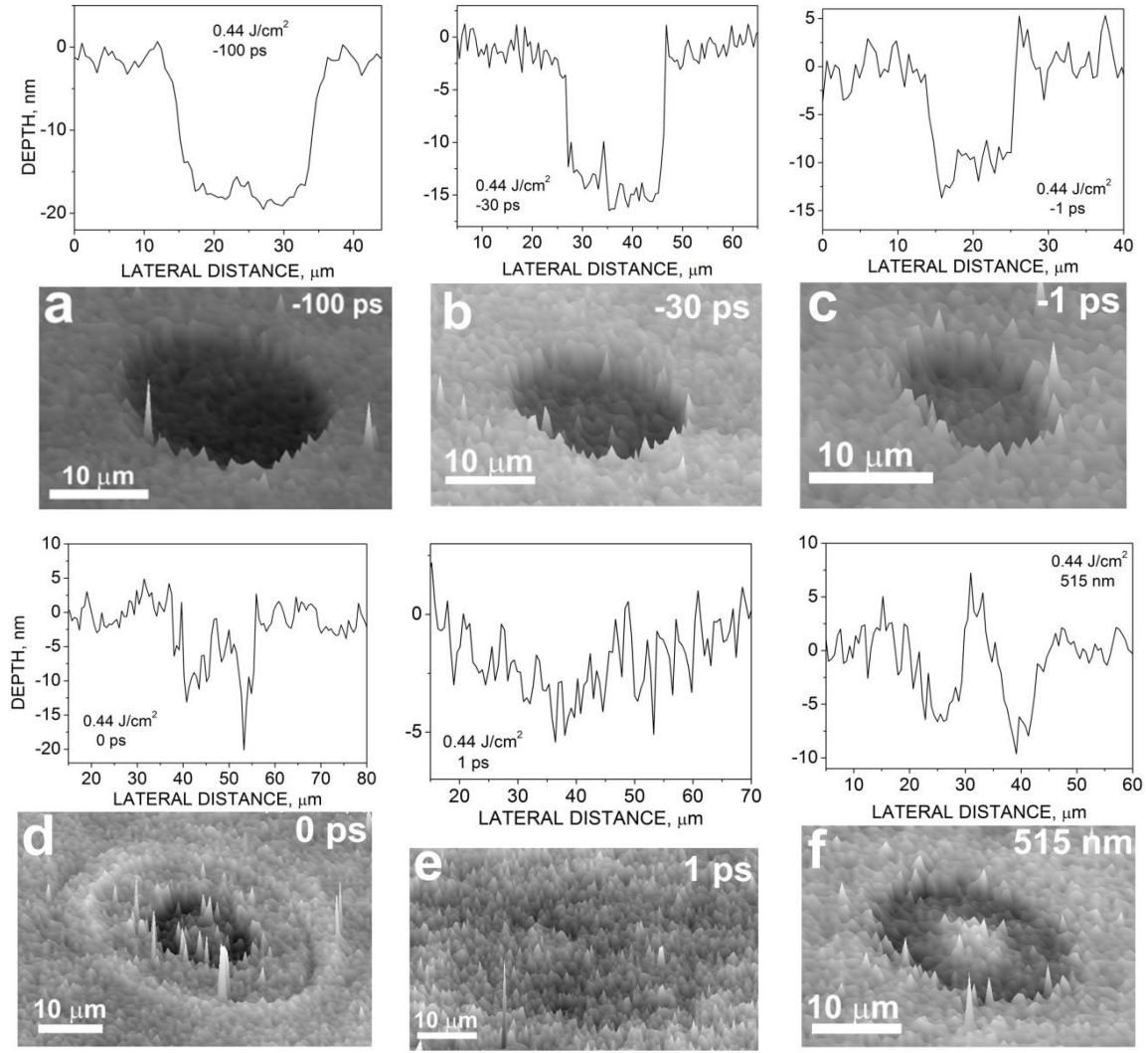

Figure 3. Depth profiles and corresponding 3D images of craters produced by two laser pulses at 515 and 1030 nm with equal fluences of 220 mJ/cm$^2$ at different time delays between the pulses (*a-e*) and by a monochromatic pulse at 515 nm at a twofold fluence of 440 mJ/cm$^2$ (*f*).

Figure 4 shows calculated time distributions of the density and temperature of free electrons generated on the silicon surface in dual-wavelength regimes at different sequence orders of two 515- and 1030-nm laser pulses and in the monochromatic regime with only 515-nm pulse of double energy. In the bi-color regime, when a visible sub-threshold pulse comes first, it generates a quite dense solid plasma (electron density $n_e \sim 1.5 \times 10^{21}$ cm$^{-3}$ under these conditions) which leads to efficient absorption of the laser energy from the following IR pulse resulting in the maximum $n_e$ of $\sim 3.8 \times 10^{21}$ cm$^{-3}$ (Fig. 4(a), green line). On the contrary, when the IR pulse comes first to the surface, the generated free-electron density is lower by more than an order of magnitude ($n_e \sim 8 \times 10^{19}$ cm$^{-3}$). Although the following 515-nm pulse excites more efficiently the valence electrons to the conduction band, it is absorbed by the electron plasma in the inverse bremsstrahlung process with much lower efficiency resulting in a twice lower $n_e$ value of ~$1.9\times10^{21}$ cm$^{-3}$ (Fig. 4(a), red line). As a result, the electron temperature $T_e$ is much higher in the case of the negative delay (visible pulse comes first). In this case, the $T_e$ value is only ~18700 K at the end of the first pulse and swiftly increases during the action of the following IR pulse up to ~57600 K (Fig. 4(b), green line). For the positive delay, the first IR pulse yields the maximal $T_e$ ~ 26800 K after which, during the action of the visible pulse, $T_e$ is only decreasing (Fig. 4(b), red line) while the conduction electrons are generated by one-photon absorption (Fig. 4(a), red line). For comparison, the action of a single visible pulse of double energy yields the maximal $n_e$ ~$2.6\times10^{21}$ cm$^{-3}$ (Fig. 4(a), black line) with the maximal temperature even slightly lower than that for the dual-wavelength regime with the positive delay (cf. red and black lines in Fig. 4(b)). This comparison emphasizes the importance of the inverse bremsstrahlung and the associated avalanche process for IR laser irradiation, which are not pronounced in the case of visible light.

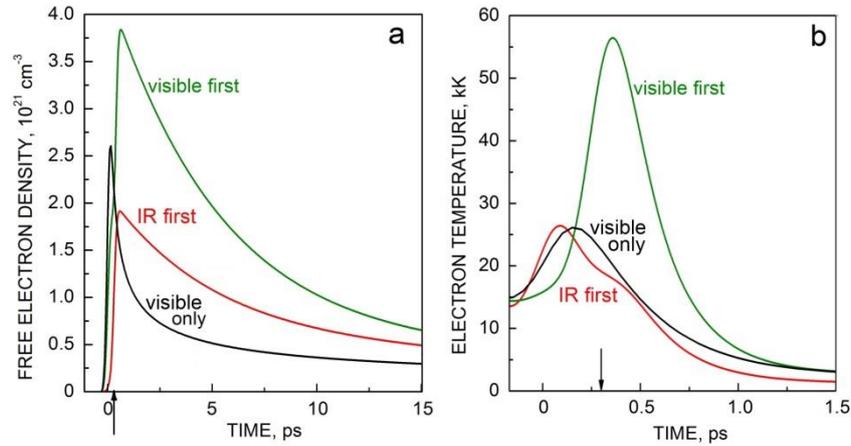

Figure 4. Calculated time evolution of the free electron density (*a*) and electron temperature (*b*) of silicon irradiated by a single 515-nm laser pulse at a fluence of 142 mJ/cm$^2$ (slightly above the melting threshold) and by two laser pulses at 515 and 1030-nm with equal energies at the same total fluence for +0.3 ps and -0.3 ps time delays between the pulses. The time = 0 corresponds to the peak of the first pulse. The arrows at the time axis show the moment of 300 fs when the second pulse is applied.

## 5. CONCLUSIONS

It is demonstrated that dual-wavelength ultrashort ablation of silicon involving visible and IR pulses can strongly enhance the laser energy coupling to the target resulting in a lower damage threshold and a higher material removal rate. It is found that the most favorable dual-wavelength ablation conditions are realized at fairly long time separations in the range of 30–100 ps between the pulses with the visible pulse coming first. In this case, the obtained craters are deepest and have a smooth cylindrical shape unachievable at single-wavelength irradiation. The effect is explained by the strong decrease in the reflectivity and the formation of a dark area at the spot center in a ~10-100-ps timescale after the action of the first visible pulse due to the expansion of a laser-generated hot liquid layer. Theoretical simulations based on the two-temperature model for bi-color irradiation conditions have revealed that the light absorption and the developing avalanche process, which both are efficient for IR laser light, are responsible for enhancing the laser energy coupling to the material, provided that the conduction band electrons are generated by a shorter wavelength pre-pulse. The results presented in this study open a new avenue for enhancing the efficiency of laser processing with combinations of different wavelengths.

## ACKNOWLEDGEMENTS

This work was supported by OPJAK financed by ESIF and the Czech MEYS, Project SENDISO: CZ.02.01.01/00/22_008/0004596.

# REFERENCES

[1] Tan, B., Venkatkrishnan, K., Sivakumar, N.R. and Gan, G.K. "Laser drilling of thick materials using femtosecond pulse with a focus of dual-frequency beam," Opt. Laser Technol. 35, 199-202 (2003). https://doi.org/10.1016/S0030-3992(02)00172-X.

[2] Zoppel, S., Merz, R., Zehetner, J. and Reider, G.A. "Enhancement of laser ablation yield by two color excitation," Appl. Phys. A 81, 847-850 (2005). https://doi.org/ 10.1007/s00339-005-3275-4

[3] Zhao, W., Wang, W., Mei, X., Jiang, G. and Liu, B. "Investigations of morphological features of picosecond dual-wavelength laser ablation of stainless steel," Opt. Laser Technol. 58, 94-99 (2014). https://doi.org/10.1016/j.optlastec.2013.11.004

[4] Bulgakov, A.V., Sládek, J., Hrabovský, J., Mirza, I., Marine, W. and Bulgakova, N.M. "Dual-wavelength femtosecond laser-induced single-shot damage and ablation of silicon," Appl. Surf. Sci. 643, 158626 (2024). https://doi.org/10.1016/j.apsusc.2023.158626

[5] Afanasiev, A., Bredikhin, V., Pikulin, A., Ilyakov, I., Shishkin, B., Akhmedzhanov, R. and Bityurin, N. "Two-color beam improvement of the colloidal particle lens array assisted surface nanostructuring," Appl. Phys. Lett. 106, 183102 (2015). https://doi.org/10.1063/1.4919898

[6] Höhm, S., Herzlieb, M., Rosenfeld, A., Krüger, J. and Bonse, J. "Dynamics of the formation of laser-induced periodic surface structures (LIPSS) upon femtosecond two-color double-pulse irradiation of metals, semiconductors, and dielectrics," Appl. Surf. Sci. 374, 331-338 (2016). https://doi.org/10.1016/j.apsusc.2015.12.129

[7] Gedvilas, M., Mikšys, J. and Račiukaitis, G. "Flexible periodical micro- and nano-structuring of a stainless steel surface using dual-wavelength double-pulse picosecond laser irradiation," RSC Adv. 5, 75075-75080 (2015). https://doi.org/10.1039/c5ra14210e

[8] Hashida, M., Furukawa, Y., Inoue, S., Sakabe, S., Masuno, S., Kusaba, M., Sakagami, H. and Tsukamoto, M. "Uniform LIPSS on titanium irradiated by two-color double-pulse beam of femtosecond laser," J. Laser Appl. 32, 022054 (2020). https://doi.org/10.2351/7.0000105

[9] Duchateau, G., Yamada, A. and Yabana, K. "Electron dynamics in α-quartz induced by two-color 10-femtosecond laser pulses," Phys. Rev. B 105, 165128 (2022). https://doi.org/10.1103/PhysRevB.105.165128

[10] Sakamoto, M., Tachikawa, T. and Fujitsuka, M. "Two-color two-laser fabrication of gold nanoparticles in a PVA film," Chem. Phys. Lett. 420. 90-94 (2006). http://dx.doi.org/10.1016/j.cplett.2005.12.053

[11] Guizard, S., Klimentov, S., Mouskeftaras, A., Fedorov, N., Geoffroy, G. and Vilmart, G. "Ultrafast breakdown of dielectrics: Energy absorption mechanisms investigated by double pulse experiments," Appl. Surf. Sci. 336, 206-211 (2015). http://dx.doi.org/10.1016/j.apsusc.2014.11.036

[12] Bulgakova, N.M., Zhukov, V.P., Collins, A.R., D. Rostohar, Derrien, T.J.-Y. and Mocek, T. "How to optimize ultrashort pulse laser interaction with glass surfaces in cutting regimes?," Appl. Surf. Sci. 336, 364-374 (2015). http://dx.doi.org/10.1016/j.apsusc.2014.12.142

[13] Gaudfrin, K., Lopez, J., Gemini, L., Delaigue, M., Hönninger, C., Kling, R. and Duchateau, G. "Fused silica ablation by double ultrashort laser pulses with dual wavelength and variable delays," Opt. Express 30, 40120-40135 (2022). https://doi.org/10.1364/OE.461502

[14] Bulgakova, N.M., Stoian, R., Rosenfeld, A., Hertel, I.V., Marine, W. and Campbell, E.E.B. "A general continuum approach to describe fast electronic transport in pulsed laser irradiated materials: The problem of Coulomb explosion," Appl. Phys. A 81, 345-356 (2005). https://doi.org/10.1007/s00339-005-3242-0

[15] van Driel, H. M. Kinetics of high-density plasmas generated in Si by 1.06- and 0.53-pm picosecond laser pulses, Phys. Rev. B 15, 8166-8176 (1987). https://doi.org/10.1103/PhysRevB.35.8166

[16] Jellison Jr., E. and Lowndes, D.H. "Optical absorption coefficient of silicon at 1.152 μ at elevated temperatures," Appl. Phys. Lett. 41, 594-596 (1982). https://doi.org/10.1063/1.93621

[17] Korfiatis, D.P., Thoma, K.-A.T. and Vardaxoglou, J.C. "Conditions for femtosecond laser melting of silicon," J. Phys. D: Appl. Phys. 40, 6803-6808 (2007). https://doi.org/10.1088/0022-3727/40/21/047

[18] Rämer, A., Osmani, O. and Rethfeld, B. "Laser damage in silicon: Energy absorption, relaxation, and transport," J. Appl. Phys. 116, 053508 (2014). https://doi.org/10.1063/1.4891633

[19] Derrien, T.J.-Y. and Bulgakova, N. M. "Modeling of silicon in femtosecond laser-induced modification regimes: accounting for ambipolar diffusion," Proc. SPIE 10228, 102280E (2027). https://doi.org/10.1117/12.2265671G

[20] Ionin, A.A., Kudryashov, S.I., Seleznev, L.V., Sinitsyn, D.V., Bunkin, A.F., Lednev, V.N. and Pershin, S.M. "Thermal melting and ablation of silicon by femtosecond laser radiation," J. Exper. Theor. Phys. 116, 347-362 (2013). https://doi.org/10.1134/S106377611302012X

[21] Bonse, J., Baudach, S., Krüger, J., Kautek, W. and Lenzner, M. "Femtosecond laser ablation of silicon-modification thresholds and morphology," Appl. Phys. A 74, 19-25 (2002). https://doi.org/10.1007/s003390100893

[22] Werner, K., Gruzdev, V., Talisa, N., Kafka, K., Austin, D., Liebig, C.M. and Chowdhury, E. "Single-shot multi-stage damage and ablation of silicon by femtosecond mid-infrared laser pulses," Sci. Rep. 9, 19993 (2019). https://doi.org/10.1038/s41598-019-56384-0

[23] Sokolowski-Tinten, K., Bialkowski, J., Cavalleri, A. and von der Linde, D. "Observation of a transient insulating phase of metals and semiconductors during short-pulse laser ablation," Appl. Surf. Sci. 127-129, 755-760 (1998). https://doi.org/10.1016/S0169-4332(97)00736-8

[24] von der Linde, D. and Sokolowski-Tinten, K. "The physical mechanisms of short-pulse laser ablation," Appl. Surf. Sci. 154-155, 1-10 (2000). https://doi.org/10.1016/S0169-4332(99)00440-7